\documentclass[aps,prd,floats,nofootinbib,preprintnumbers]{revtex4}
\usepackage{epsfig}
\usepackage{amsmath}
\usepackage{amssymb}
\usepackage{amsthm}
\usepackage{longtable}

\begin{document}

\preprint{NUHEP-TH/10-08}

\title{Parameterizing Majorana Neutrino Couplings in the Higgs Sector}

\author{Andr\'e de Gouv\^ea}
\affiliation{Northwestern University, Department of Physics \&
Astronomy, 2145 Sheridan Road, Evanston, IL~60208, USA}

\author{Wei-Chih Huang}
\affiliation{Northwestern University, Department of Physics \&
Astronomy, 2145 Sheridan Road, Evanston, IL~60208, USA}

\author{Shashank Shalgar}
\affiliation{Northwestern University, Department of Physics \&
Astronomy, 2145 Sheridan Road, Evanston, IL~60208, USA}

\begin{abstract}
Nonzero masses for the active neutrinos -- regardless of their nature or origin --  arise only after electroweak symmetry breaking. We discuss the parameterization of neutrino couplings to a Higgs sector consisting of one $SU(2)_L$ scalar doublet and one $SU(2)_L$ scalar triplet, and allow for right-handed neutrinos whose Majorana mass parameters arise from the vacuum expectation value of a Standard Model scalar singlet. If the neutrinos are Majorana fermions, all Yukawa couplings can be expressed as functions of the neutrino mass eigenvalues and a subset of the elements of the neutrino mixing matrix. In the mass basis, the Yukawa couplings are, in general, not diagonal. This is to be contrasted to the case of charged-fermions or Dirac neutrinos, where couplings to the Higgs-boson are diagonal in the mass basis and proportional only to the fermion masses. Nonetheless, all physically distinguishable parameters can be reached if all neutrino masses are constrained to be positive, all mixing angles constrained to lie in the first quadrant ($\theta\in[0,\pi/2]$), and all Majorana phases to lie in the first two quadrants ($\phi\in[0,\pi]$), as long as all Dirac phases vary within the entire unit circle ($\delta\in[0,2\pi\}$). We discuss several concrete examples and comment on the Casas-Ibarra parameterization for the neutrino Yukawa couplings in the case of the type-I Seesaw Lagrangian.
\end{abstract}

\maketitle

\setcounter{equation}{0} \setcounter{footnote}{0}
\section{Introduction}
\label{sec:Intro}

While it is experimentally established that neutrino masses are not zero (see \cite{NeutrinoReview} for some overviews), the mechanism behind them is unknown. Several distinct ideas have been pursued over the last few decades and the hope is that, ultimately, experiments will equip us with enough information to figure out which one, if any, is correct. 

The gauge quantum numbers of all fermion fields in the standard model are such that relevant fermion mass operators are forbidden by gauge invariance, and fermion masses arise only after electroweak symmetry breaking. In the case of charged fermions, masses are a consequence of Yukawa couplings between different chiral fermions and the $SU(2)_L$ doublet Higgs scalar field. When this Higgs field acquires an expectation value breaking electroweak symmetry, pairs of chiral fermions combine and acquire Dirac masses. Assuming that this scenario is correct, not only are the fermions masses generated but one is able to make two well-defined predictions: at leading order, the couplings between the propagating Higgs boson and the different fermions with a well-defined mass are (i) diagonal and (ii) unambiguously determined by the fermion masses. For example, 
\begin{equation}
{\cal L}_{\rm SM}\supset -\lambda^e_{\alpha} L_{\alpha}e^c_{\alpha}\tilde{H}+H.c.,
\end{equation}
where $L_{\alpha}=(\nu_{\alpha},~e_{\alpha})^T$ are the lepton doublet fields, $e^c_{\alpha}$ are the (anti)lepton doublet fields and $H$ is the Higgs doublet scalar field, and $\lambda^e_{\alpha}$ are dimensionless Yukawa couplings. We choose a weak basis for the leptons such that the Yukawa interactions are diagonal and $\alpha=e,\mu,\tau$. When the Higgs field acquires a vacuum expectation value $(0,~v/\sqrt{2})^T$, $v=246$~GeV, the three charged fermions acquire masses $m_{\alpha}=\lambda^e_{\alpha}v/\sqrt{2}$ and the couplings between the charged leptons and the propagating Higgs boson $h^0$ are $m_{\alpha}/v$, i.e., after electroweak symmetry breaking,
\begin{equation}
{\cal L}_{\rm SM}\supset -\frac{h}{v}\left(m_e ee^c+m_\mu\mu\mu^c+m_\tau\tau\tau^c\right)+H.c..
\label{eq:cc_Higgs}
\end{equation}

Masses for the active neutrinos (those that couple to the $W$ and $Z$ gauge bosons, $\nu_a=\nu_e,\nu_{\mu}, \nu_{\tau}$), regardless of how they are generated, must also arise as a consequence of electroweak symmetry breaking.  Furthermore, new degrees of freedom must be added to the Lagrangian in order to allow neutrino masses. For example, if one adds gauge singlet (anti)leptons $n_i$ ($i=1,2,\ldots$)  to the Standard Model particle content and only considers the new Yukawa interactions
\begin{equation}
{\cal L}_{D}\supset -\lambda^{\nu}_{\alpha i} L_{\alpha}n_iH+H.c.,
\label{eq:Dirac_nus}
\end{equation}
neutrinos also acquire Dirac masses and, as above, their couplings to the Higgs boson are diagonal in the mass-eigenstate basis and proportional to the neutrino mass eigenvalues, which in turn are proportional the square-root of the eigenvalues of the square of the Yukawa coupling matrix, $\lambda^{\nu}(\lambda^{\nu})^{\dagger}$, as in the case of the charged fermions.

Life, however, can be more interesting. Since neutrinos are singlets of the unbroken $U(1)_{\rm em}$ gauge symmetry, they are, unlike all charged fermions, allowed to acquire Majorana masses. These may  arise from several distinct $SU(2)_L\times U(1)_Y$ gauge invariant Lagrangians, which will be discussed in more detail in Sec.~\ref{sec:Lagrangian}. If that is the case, the couplings between the neutrinos and the electroweak symmetry breaking sector (neutrino--Higgs sector) need not be diagonal in the mass basis. It turns out, however, that the couplings are not independent parameters, even if one allows for a very generic neutrino--Higgs sector. They are uniquely determined by the neutrino mass eigenvalues and the elements of the lepton mixing matrix $U$. This statement is true as long as one takes into account all light degrees of freedom, which may include several of the gauge singlet fermions,  as we review in Sec.~\ref{sec:PhyRange}.

Here we further pursue the connection between the leptonic mixing matrix, neutrino masses, and the couplings of the neutrino--Higgs sector, concentrating on two issues, both related to how to properly parameterize the leptonic mixing angles taking into account (i) that the neutrinos are Majorana fermions and (ii) that there may be light gauge singlet fermions (sterile neutrinos). In the effective theory below electroweak symmetry breaking where all relevant degrees of freedom are the neutrinos, the charged leptons, and the electroweak gauge bosons, the lepton sector is parameterized by 3 charged lepton masses, $3+N$ neutrino masses, where $N$ is the number of light sterile neutrinos,  and $6+6N$ real parameters in the mixing matrix $U$. Of those $6+6N$ real numbers, half can be parameterized as mixing angles $\theta$ and the other half as CP-odd phases. Of the CP-odd phases, $2+N$ will be referred to as Majorana phases $\phi$ and the remaining $1+2N$ as Dirac phases $\delta$. See, for example, \cite{Branco:1999fs} for details and a pedagogical discussion of some of the relevant issues. It was shown in \cite{deGouvea:2008nm} that all mixing angles can be chosen in the first quadrant, i.e., $\theta\in[0,\pi/2]$, all Majorana phases in the first two quadrants, i.e., $\phi\in[0,\pi]$, while all Dirac phases must be allowed to vary within the whole unit circle in order to cover all physically distinguishable possibilities, i.e., $\delta\in[0,2\pi\}$. For other discussions of this interesting issue from different points of view see also \cite{Latimer:2004hd,Jenkins:2007ip}. See \cite{Fogli:1996ne,de Gouvea:2000cq} for earlier discussions.

Both the parameter counting and the allowed ranges above apply below the electroweak symmetry breaking scale and in the absence of new interactions. For example, the number of physical parameters in the mixing matrix depends on the fact that the sterile degrees of freedom are sterile so that one is free to perform ``sterile--only'' rotations indiscriminately. We extend the analysis in \cite{deGouvea:2008nm} to include the neutrino--Higgs interactions and ask whether an extended physical range, or new mixing parameters, are necessary. Naively, it is easy to see why the answer might be `yes' since Yukawa couplings of different kinds may qualify as new interactions. For example, if Majorana masses for the singlet fermions $n_i$ arise from Yukawa interactions with a gauge singlet scalar $S$ that acquires a vacuum expectation value (operators of the type $y_{ij}n_in_jS$ so $M_{ij}\propto y_{ij}\langle S\rangle$) one should question whether ``sterile--only'' rotations can be performed with impunity. We show in Sec.~\ref{sec:PhyRange} that the results of  \cite{deGouvea:2008nm} apply even when one takes interactions in the neutrino--Higgs sector into account and describe several examples in Sec.~\ref{sec:Examples}.

Our discussion will be mostly directed towards the formalism of neutrino masses and mixing with less concern toward more practical issues. We will, for example, not discuss whether some of the processes analyzed here can be observed in practice or how one may go about measuring some of the hard-to-get-to masses and mixing angles. On the other hand, neutrino Yukawa couplings may play a significant role in the real world if Leptogenesis is responsible for the baryon-asymmetry of the Universe. For this reason we describe in Sec.~\ref{sec:Casas} the well-known Casas--Ibarra parameterization of the neutrino Yukawa couplings \cite{Casas:2001sr}  and comment on how it relates to the parameterization of the neutrino--Higgs sector presented here. 

\setcounter{equation}{0} \setcounter{footnote}{0}
\section{Neutrino Higgs Sectors}
\label{sec:Lagrangian}

We will restrict ourselves to scenarios where the neutrinos end up as Majorana fermions and discuss only renormalizable Lagrangians that yield massive neutrinos. Unless otherwise noted, all fermion fields will be treated as Weyl spinors and we will assume three active lepton generations and $N$ gauge singlet fermions sometimes referred to as sterile neutrinos or right-handed neutrinos. When discussing couplings between different Higgs fields and the leptons, we will always express the leptons in the mass eigenstate basis unless otherwise noted.\footnote{In order to consider processes involving neutrinos with a well-defined flavor, we simply have to add the different amplitudes for neutrinos with a well-defined mass with the appropriate weights, taking the coherence of the different neutrino mass eigenstates in account when applicable.}

Neutrino Majorana masses can arise in a variety of ways. As mentioned in the introduction, the non-trivial $SU(2)_L\times U(1)_Y$ quantum numbers of the active neutrinos mean that they can only acquire masses -- of any kind -- after electroweak symmetry is broken and only via interactions to the physics responsible for electroweak symmetry breaking, referred to here as the Higgs sector. Other than the Yukawa couplings described in  Eq.~(\ref{eq:Dirac_nus}), neutrino masses will also arise in the presence of an $SU(2)_L$ Higgs triplet $T$ with hypercharge $+1$ via the Yukawa interaction
\begin{equation}
{\cal L}_{T}\supset -\frac{g_{\alpha\beta}}{2}L_{\alpha}L_{\beta}T+H.c.,
\label{eq:triplet}
\end{equation}
where $g$ are dimensionless Yukawa couplings and $g_{\alpha\beta}=g_{\beta\alpha}$. If, for example, the neutral component of $T$ has a vacuum expectation value\footnote{In this case, one must inquiry whether there are Goldstone boson degrees of freedom associtated to the potential spontaneous breaking of lepton number. This issue does not concern us here but can be circumvented, for example, by including scalar potential terms that explicitly violate  lepton number (see, for example, \cite{Ma:1998dx}) .} $u$, the active neutrinos acquire a Majorana mass matrix $m^{a}=g u$ \cite{Schechter:1980gr,Gelmini:1980re,Ma:1998dx}. In this scenario, neutrino Majorana masses arise even in the absence of right-handed neutrinos $n_i$. Small neutrino masses imply that either the $g$ couplings are very small ($g\ll1$) or that $u$ is much smaller than the electroweak symmetry breaking scale, $u\ll v$. Electroweak precision data independently require $u$ to be small, $u\ll v$ (for detailed analyses see, for example, \cite{Chen:2006pb,Chankowski:2006hs,Akeroyd:2010je} and references therein). 

Gauge singlet fermions, if present, are not constrained to be massless in the absence of electroweak symmetry breaking. These are allowed to have relevant  ``right-handed neutrino'' Majorana masses 
\begin{equation}
{\cal L}_{\rm M_R}\supset \frac{M^R_{ij}}{2}n_in_j+H.c.,
\label{eq:MR}
\end{equation}
where $M^R$ are mass parameters and $M^R_{ij}=M^R_{ji}$. Eq.~(\ref{eq:MR}), combined with Eq.~(\ref{eq:Dirac_nus}), leads, after electroweak symmetry breaking, to massive neutral fermions which contain the three active neutrino degrees of freedom. It is easy to see that lepton number is explicitly violated and that all massive neutrinos are Majorana fermions for generic values of $\lambda^{\nu}$ and $M^R$. ${\cal L}_{SM}+{\cal L}_D+{\cal L}_{M_R}$ consists of the most general renormalizable Lagrangian given the existence of gauge singlet fermions, and is by far the most popular model for generating neutrino masses \cite{SeeSaw}. The relevant $M^R$ parameters are, in general, quite unrelated to the phenomenon of electroweak symmetry breaking and are, experimentally, virtually unconstrained \cite{deGouvea:2005er,Kusenko:2004qc,Smirnov:2006bu,deGouvea:2006gz,de Gouvea:2007uz,Gorbunov:2007ak,Atre:2009rg,deGouvea:2009fp}. 

It is sometimes the case that $M^R$ is a consequence of spontaneous symmetry breaking (this is the case, for example, in left-right models). Here, we will simply consider that it arises from the Yukawa coupling between the $n_i$ fields and a gauge singlet scalar field $S$, 
\begin{equation}
{\cal L}_{\rm S}\supset \frac{y_{ij}}{2}n_in_jS+H.c.,
\label{eq:S}
\end{equation}
so $M^R=ys$, where $s=\langle S\rangle$ and $M^R$ is defined in Eq.~(\ref{eq:MR}). As in the triplet case, Eq.~(\ref{eq:triplet}), we will not worry about the origin of $\langle S\rangle$ or the existence of potentially dangerous Goldstone bosons.   

In general, we will concern ourselves with the most general Lagrangian ${\cal L}_{SM}+{\cal L}_D+{\cal L}_{S}+{\cal L}_T$ and the parameterization of the Yukawa couplings $\lambda^{\nu}$, $g$, and $y$ which couple the neutrinos (and sometimes the charged leptons) to different scalar fields in $T$, $S$, and $H$. These are intimately connected to the neutrino masses and the mixing angles, as we discuss in detail in the next section.

We have restricted our models to renormalizable Lagrangians, but could have extended it to include irrelevant operators as well. For example, one could consider the possibility that some new very heavy physics, when integrated out, led to the well-known Weinberg operator \cite{Weinberg:1979sa}:
\begin{equation}
{\cal L}_{\rm 5}\supset \frac{g'_{\alpha\beta}}{2\Lambda}L_{\alpha}HL_{\beta}H + H.c.,
\label{eq:5d}
\end{equation} 
in the flavor basis. Here $g'$ are dimensionless couplings, and $\Lambda$ is a free parameter with dimensions of mass. Upon electroweak symmetry breaking, this leads to an ``active'' neutrino mixing matrix (as in the triplet Higgs boson case) $m^a=g'v^2/\Lambda$. In this case, we can also parameterize the neutrino--Higgs boson couplings $g'$ in terms of masses and mixing angles. All results below that concern the triplet Higgs Yukawa coupling $g$ can be applied upon proper reinterpretation to the $g'$ couplings above. One need only make sure to stick to the proper effective theory. For example, in the ultraviolet, the theory may contain two light and two very heavy right-handed neutrinos, both coupling to the lepton doublets via Eq.~(\ref{eq:Dirac_nus}). Upon integrating out the two heavy degrees of freedom the operator Eq.~(\ref{eq:5d}) is generated. The Lagrangian in this case would consist of three active plus two sterile neutrinos which couple via Eq.~(\ref{eq:Dirac_nus}), plus  Eq.~(\ref{eq:5d}).

\setcounter{equation}{0}
\section{Parameters and Physical Ranges}
\label{sec:PhyRange}

The $(3+N)\times(3+N)$ neutrino mass matrix $m^\nu$ is symmetric ($m^{\nu}_{\alpha\beta}=m^{\nu}_{\beta\alpha}$) and can be written as 
\begin{equation}
    m^{\nu}=\left(
        \begin{array}{cc}
          (m^a)_{3\times 3} & (m^D)_{3\times N} \\
          (m^D)^T_{N\times 3} & (M^R)_{N\times N} \\
        \end{array}
      \right) =
      \left(
        \begin{array}{cc}
          (gu)_{3\times 3} & (\lambda^{\nu}v)_{3\times N} \\
          (\lambda^{\nu}v)^T_{N\times 3} & (ys)_{N\times N} \\
        \end{array}
      \right),
        \label{eq:general}
\end{equation}
where we explicitly indicated the dimensionality of the different sub-matrices. We will make use of this sub-matrix notation henceforth without indicating the dimensionality of the different parts whenever unambiguous in order to avoid an index overload. As a symmetric matrix, $m^{\nu}$ can be diagonalized 
\begin{equation}
m^{\nu}=U^*m^{\nu}_{\rm Diag}U^{\dagger},
\label{eq:mnu_diag}
\end{equation}
where $m^{\nu}_{\rm Diag}$ is a diagonal matrix with non-negative diagonal entries and $U$ is a unitary matrix. In the weak basis where the charged-lepton mass matrix is diagonal along with the charged current $W\ell\nu$ couplings, $U$ is directly related to the neutrino mixing matrix that connects neutrino mass eigenstates to neutrino flavor eigenstates:
\begin{equation}
 \nu_{\alpha}=U_{\alpha i}\nu_i,
 \label{eq:Udef}
\end{equation}
where $\alpha=e,\mu,\tau,s_1,s_2,\ldots$ and $i=1,2,3,4,5,\ldots$. We will stick to this weak basis unless otherwise noted. Gauge boson interactions are parameterized by the gauge couplings and the elements of $U$. For example, the $W$-boson coupling to a muon and a $\nu_4$ is proportional to $U_{\mu 4}$. Similarly, the $Z$-boson couplings to the different mass-eigenstates are proportional to the elements of $U$ and are not necessarily trivial if $N\ge 1$ \cite{Schechter:1980gr}. For example, the coupling of the $Z$-boson to a pair of $\nu_1$'s is proportional to 
\begin{equation}
\sum_{\alpha=e,\mu,\tau}U_{\alpha 1}U^*_{\alpha 1}\neq 1~(\mbox{in general}).
\end{equation}
Note that the sum above is restricted to the active neutrino flavors. 

It will prove convenient to extract the Yukawa couplings by ``chopping-off'' different parts of $m^{\nu}$. In detail,
\begin{eqnarray}
 \left(\begin{array}{cc} g & 0 \\ 0 & 0 \\ \end{array}\right) &=& 
 \frac{1}{u}
 \left(\begin{array}{cc}1 & 0 \\0 & 0 \\\end{array} \right) 
 m^{\nu}
 \left(\begin{array}{cc} 1 & 0 \\0 & 0 \\\end{array}\right),
 \\
 \left(\begin{array}{cc} 0 & 0 \\ 0 & y \\\end{array}\right) &=& 
 \frac{1}{s}
 \left(\begin{array}{cc}0 & 0 \\0 & 1 \\\end{array} \right) 
 m^{\nu}
 \left(\begin{array}{cc} 0 & 0 \\0 & 1 \\\end{array}\right),
 \\
  \left(\begin{array}{cc} 0 & \lambda \\ \lambda^T & 0 \\ \end{array}\right) &=& 
  \frac{1}{v}
  \left[
  \left(\begin{array}{cc}1 & 0 \\ 0 & 0 \\ \end{array} \right) 
  m^{\nu}
  \left(\begin{array}{cc} 0 & 0 \\ 0 & 1 \\ \end{array} \right) + 
  \left(\begin{array}{cc} 0 & 0 \\ 0 & 1 \\ \end{array} \right) 
  m^{\nu}
  \left(\begin{array}{cc}1 & 0 \\ 0 & 0 \\ \end{array} \right)
 \right],
 \end{eqnarray}
where $0$ and $1$ stand for, respectively, null matrices and identity matrices, respectively. The proper dimensionalities of each sub-matrix can be inferred from Eq.~(\ref{eq:general}). Using Eqs.~(\ref{eq:mnu_diag},\ref{eq:Udef}) 
 \begin{eqnarray}
 \nu_i g_{ij} \nu_j &=& 
 \frac{1}{u}\left(\begin{array}{ccc}\nu_1 & \ldots & \nu_{3+N} \end{array}\right) U^T
 \left(\begin{array}{cc}1 & 0 \\0 & 0 \\\end{array} \right) 
 U^*  \left(
        \begin{array}{ccc}
          m_1 & 0 & 0 \\
          0 & \ddots & 0 \\
          0 & 0 & m_{3+N} \\
        \end{array}
      \right) U^{\dagger}
 \left(\begin{array}{cc} 1 & 0 \\0 & 0 \\\end{array}\right)
 U\left(\begin{array}{c}\nu_1 \\ \vdots \\ \nu_{3+N} \end{array}\right), \label{g}
 \\
 \nu_i y_{ij} \nu_j &=& 
 \frac{1}{s}\left(\begin{array}{ccc}\nu_1 & \ldots & \nu_{3+N} \end{array}\right) U^T
 \left(\begin{array}{cc}0 & 0 \\0 & 1 \\\end{array} \right) 
U^* \left(
        \begin{array}{ccc}
          m_1 & 0 & 0 \\
          0 & \ddots & 0 \\
          0 & 0 & m_{3+N} \\
        \end{array}
      \right) U^{\dagger}
 \left(\begin{array}{cc} 0 & 0 \\0 & 1 \\\end{array}\right)
 U\left(\begin{array}{c}\nu_1 \\ \vdots \\ \nu_{3+N} \end{array}\right), \label{y}
 \\
  \nu_i\lambda^{\nu}_{ij}\nu_j &=& 
  \frac{1}{v}\left(\begin{array}{ccc}\nu_1 & \ldots & \nu_{3+N} \end{array}\right) U^T
  \left[
  \left(\begin{array}{cc}1 & 0 \\ 0 & 0 \\ \end{array} \right) 
  U^* \left(
        \begin{array}{ccc}
          m_1 & 0 & 0 \\
          0 & \ddots & 0 \\
          0 & 0 & m_{3+N} \\
        \end{array}
      \right) U^{\dagger}
  \left(\begin{array}{cc} 0 & 0 \\ 0 & 1 \\ \end{array} \right) + \right. \nonumber
  \\ && + \left.
  \left(\begin{array}{cc} 0 & 0 \\ 0 & 1 \\ \end{array} \right) 
  U^* \left(
        \begin{array}{ccc}
          m_1 & 0 & 0 \\
          0 & \ddots & 0 \\
          0 & 0 & m_{3+N} \\
        \end{array}
      \right) U^{\dagger}
  \left(\begin{array}{cc}1 & 0 \\ 0 & 0 \\ \end{array} \right)
 \right]U\left(\begin{array}{c}\nu_1 \\ \vdots \\ \nu_{3+N} \end{array}\right). \label{lambda}
 \end{eqnarray}
The above expressions are the generalization of the familiar charged fermion--Higgs-boson couplings, e.g. Eq.~(\ref{eq:cc_Higgs}) in the case of charged leptons. Here, however, while the Yukawa couplings are unambiguously defined\footnote{The magnitude of the different couplings depends on three potentially unrelated energy scales, $u,v,s$. We will have nothing to say about these other than $v=246$~GeV while the other two are only poorly constrained.} from the parameters in the neutrino mass matrix, they depend on both the neutrino mass eigenvalues $m_1, m_2, \ldots, m_{3+N}$ and on the elements of the neutrino mixing matrix, $U$. They are, unlike the charged-fermion case, off-diagonal in the mass basis. Note that in Eqs.~(\ref{g},\ref{y},\ref{lambda}) all Yukawa couplings are symmetric $(3+N)\times(3+N)$ matrices and that, generically, all elements are nonzero.

Two notable special cases are the Dirac neutrino case ($y=g=0$), discussed briefly in the introduction, and the $N=0$ case (or, similar that of $\lambda^{\nu}=0$), where the active neutrinos get their masses exclusively from the triplet Higgs-boson $g$-couplings. In this case, it is trivial to see that 
\begin{equation}
\nu_i g_{ij} \nu_j = 
 \frac{1}{u}\left(\begin{array}{ccc}\nu_1 & \nu_2 & \nu_{3} \end{array}\right) U^T 
 U^*  \left(
        \begin{array}{ccc}
          m_1 & 0 & 0 \\
          0 & m_2 & 0 \\
          0 & 0 & m_3 \\
        \end{array}
      \right) U^{\dagger}
 U\left(\begin{array}{c}\nu_1 \\ \nu_2 \\ \nu_3 \end{array}\right) = \frac{1}{u}\left(m_1\nu_1\nu_1+m_2\nu_2\nu_2+m_3\nu_3\nu_3\right).
\end{equation}

We proceed to discuss how many parameters are required to describe all neutrino--Higgs couplings, and what ranges these should cover in order to describe all physically distinguishable possibilities. 

\subsection{Sterile Neutrino Rotations}

As far as the weak interactions are concerned, a redefinition of the sterile neutrino fields is clearly unphysical, as argued, for example, in \cite{deGouvea:2008nm}. In more detail, if one redefines the neutrino flavor basis 
\begin{equation}\label{type-I R transformation in sterile sector}
    \left(\begin{array}{c}  \nu_{e} \\ \nu_{\mu} \\ \nu_{\tau} \\ \nu_{s_1} \\ \vdots \\ \nu_{s_N} \\ \end{array} \right)
    \rightarrow
    \left( \begin{array}{c} \nu_{e} \\  \nu_{\mu} \\ \nu_{\tau} \\ \nu_{s_1'} \\ \vdots \\ \nu_{s_N'} \\  \end{array} \right) =
    \left( \begin{array}{cc} 1_{3\times 3} & 0_{3\times N} \\ 0_{N\times 3} & \Omega_{N\times N} \\ \end{array} \right)
     \left( \begin{array}{c} \nu_{e} \\ \nu_{\mu} \\ \nu_{\tau} \\ \nu_{s_1} \\ \vdots \\ \nu_{s_N} \\ \end{array} \right),
    \end{equation}
where $\Omega$ is a unitary $N\times N$ matrix, the weak interactions in the mass basis are left untouched.  Hence, the two mixing matrices $U$ and $U'$ are equivalent when 
\begin{equation}
U' = \left( \begin{array}{cc} 1 & 0 \\ 0 & \Omega \\ \end{array} \right) U,
 \end{equation}
for any unitary $\Omega$. This redundancy allows one to reduce the number of physically observable mixing parameters by $N^2-N$ \cite{Branco:1999fs,deGouvea:2008nm}. It remains to show that this equivalence is respected when it comes to describing the Yukawa interactions $\lambda^{\nu}$, $g$ and $y$. This is easily done by noticing the following:
\begin{equation}
U^{\dagger}\left(\begin{array}{cc}1/0 & 0 \\ 0 & 0/1 \\ \end{array} \right)U =  
U'^{\dagger} \left( \begin{array}{cc} 1 & 0 \\ 0 & \Omega \\ \end{array} \right)\left(\begin{array}{cc}1/0 & 0 \\ 0 & 0/1 \\ \end{array} \right) \left( \begin{array}{cc} 1 & 0 \\ 0 & \Omega^{\dagger} \\ \end{array} \right) U'=
U'^{\dagger}\left(\begin{array}{cc}1/0 & 0 \\ 0 & 0/\Omega\Omega^{\dagger} \\ \end{array} \right)U',
\label{eq:invariance}
\end{equation}
where the compact notation $A/B$ indicates matrix $A$ or $B$. The same is true for the transpose of Eq.~(\ref{eq:invariance}). From trivial substitution into Eqs.~(\ref{g},\ref{y},\ref{lambda}), all $g_{ij}$,  $y_{ij}$, $\lambda^{\nu}_{ij}$ are invariant upon $U\to U'$ for any unitary $\Omega$. This invariance is often evoked to choose a weak basis where $M^R$ (and hence $y$) is diagonal and real. In general, of course, these states are not mass eigenstates. 

Analogously, the Yukawa interactions are also invariant under purely active field redefinitions, 
\begin{equation}
U'' = \left( \begin{array}{cc} \Omega^a_{3\times 3} & 0_{3\times N} \\ 0_{N\times 3} & 1_{N\times N} \\ \end{array} \right) U,
 \label{eq:Omega_a}
 \end{equation}
where $\Omega^a$ is a unitary matrix. Such field redefinitions, of course, do not leave the weak interactions invariant, Consequences of this redundancy will be discussed in Sec.~\ref{sec:Examples}. The invariance of the Yukawa couplings under these ``diagonal'' flavor neutrino redefinitions is a straightforward consequence of the sub-matrix decomposition of the different neutrino--Higgs couplings. 

\subsection{Parameter Ranges}

Finally, we discuss the physically allowed range of the parameters in $U$, namely the mixing angles $\theta$, the Majorana phases $\phi$ and the Dirac phases $\delta$. We follow the formalism developed in \cite{deGouvea:2008nm}, to which we refer for more details. There, it was shown that 
\begin{equation}
U(\theta,\phi,\delta)=P_{\alpha} U(\theta',\phi',\delta') P_i,
\end{equation}
where $P_{\alpha}$ and $P_i$ are diagonal matrices whose entries are either $+1$ or $-1$ and $\theta,\theta'$, $\phi,\phi'$ and $\delta,\delta'$ are related via specular-like reflections on the unit circle ($\theta'=-\theta$, $\theta'=\theta+\pi$, $\phi'=\phi+\pi$, etc). The fact that $P_{\alpha}$ can be absorbed by redefining the sign of the charged lepton mass eigenstates and $P_i$ can be absorbed by redefining the sign of the neutrino mass eigenstates allows one to identify different values of $\theta$, $\phi$, and $\delta$ and choose the physically inequivalent range for all mixing angles, Majorana and Dirac phases such that all $\theta\in[0,\pi/2]$, all $\phi\in[0,\pi]$ while the $\delta$ ranges cannot be constrained: $\delta\in[0,2\pi\}$. Once the Yukawa interactions are also considered, one needs to check that the equivalence used to reach this conclusion still applies.

Making use of the fact that $P_{i,\alpha}=P^{\dagger}_{i,\alpha}$ and $P_{i,\alpha}P_{i,\alpha}=1$,
\begin{equation}
U^{\dagger}(\theta,\phi,\delta)\left(\begin{array}{cc}1/0 & 0 \\ 0 & 0/1 \\ \end{array} \right)U(\theta,\phi,\delta)   
=
P_iU^{\dagger}(\theta',\phi',\delta')\left(\begin{array}{cc}1/0 & 0 \\ 0 & 0/1 \\ \end{array} \right)U(\theta',\phi',\delta')P_i,
\end{equation}
and
\begin{eqnarray}
& \hspace{-0.4cm} \left(\begin{array}{ccc}\nu_1 & \ldots & \nu_{3+N} \end{array}\right) U^T(\theta,\phi,\delta)
 \left(\begin{array}{cc}1/0 & 0 \\0 & 0/1 \\\end{array} \right) 
 U^*(\theta,\phi,\delta)  m^{\nu}_{\rm Diag} U^{\dagger}(\theta,\phi,\delta)
 \left(\begin{array}{cc} 1/0 & 0 \\0 & 0/1 \\\end{array}\right)
 U(\theta,\phi,\delta)\left(\begin{array}{c}\nu_1 \\ \vdots \\ \nu_{3+N} \end{array}\right)  = \nonumber \\
& \hspace{-0.4cm} \left(\begin{array}{ccc}\nu_1 & \ldots & \nu_{3+N} \end{array}\right) P_i U^T(\theta',\phi',\delta')
 \left(\begin{array}{cc}1/0 & 0 \\0 & 0/1 \\\end{array} \right) 
 U^*(\theta',\phi',\delta') m^{\nu}_{\rm Diag} U^{\dagger}(\theta',\phi',\delta')
 \left(\begin{array}{cc} 1/0 & 0 \\ 0 & 0/1 \\\end{array}\right)
 U(\theta',\phi',\delta') P_i\left(\begin{array}{c}\nu_1 \\ \vdots \\ \nu_{3+N} \end{array}\right). \nonumber & \\
\end{eqnarray}
Hence, the {\sl same} sign redefinition of the $\nu_i$ fields that renders $\theta,\phi,\delta$ and $\theta',\phi',\delta'$ physically equivalent as far as the weak interactions are concerned also renders $\theta,\phi,\delta$ and $\theta',\phi',\delta'$ physically equivalent as far as the Yukawa interactions are concerned.

\setcounter{equation}{0}
\section{Examples}
\label{sec:Examples}

Here we discuss a few simple scenarios to illustrate the results presented in Sec.~\ref{sec:PhyRange}. More specifically, we will write explicit expressions for the Yukawa couplings in the mass basis and will describe some observables one could measure, at least in principle, in order to access them.

\subsection{One Active, One Sterile}

In the case of only one active ($\nu_e$) and one sterile neutrino ($\nu_s$), along with one charged lepton ($e$), the $2\times 2$ neutrino mixing matrix $U$ is parameterized by one mixing angle $\theta$ and one Majorana phase $\phi$,
\begin{equation}
U=\left(\begin{array}{cc} \cos\theta & \sin\theta \\ -\sin\theta & \cos\theta \\ \end{array} \right)\cdot
     \left(\begin{array}{cc} 1 & 0 \\ 0 & e^{i \phi} \\ \end{array}\right), 
     ~~~
m^{\nu}_{\rm Diag}=\left( \begin{array}{cc} m_1 & 0 \\ 0 & m_2 \\ \end{array} \right).
\end{equation}
The $2\times 2$ Yukawa matrices, written in the mass basis, are 
\begin{eqnarray}
g &=& \frac{1}{u}\left(\begin{array}{cc} m_1\cos^4\theta + m_2e^{-2i\phi}\cos^2\theta\sin^2\theta & \cos\theta\sin\theta\left(m_1e^{i\phi}\cos^2\theta+m_2e^{-i\phi}\sin^2\theta\right) \\  \cos\theta\sin\theta\left(m_1e^{i\phi}\cos^2\theta+m_2e^{-i\phi}\sin^2\theta\right) & m_1e^{2i\phi}\cos^2\theta\sin^2\theta + m_2\sin^4\theta \end{array} \right), \\
y &=& \frac{1}{s}\left(\begin{array}{cc} m_1\sin^4\theta + m_2e^{-2i\phi}\cos^2\theta\sin^2\theta & -\cos\theta\sin\theta\left(m_1e^{i\phi}\sin^2\theta+m_2e^{-i\phi}\cos^2\theta\right) \\  -\cos\theta\sin\theta\left(m_1e^{i\phi}\sin^2\theta+m_2e^{-i\phi}\cos^2\theta\right) & m_1e^{2i\phi}\cos^2\theta\sin^2\theta + m_2\cos^4\theta \end{array} \right), \\
\lambda^{\nu} &=& \frac{1}{v}\left(\begin{array}{cc} 2\sin^2\theta\cos^2\theta\left(m_1-m_2e^{-2i\phi}\right) & -\cos\theta\sin\theta\cos2\theta\left(m_1e^{i\phi}-m_2e^{-i\phi}\right) \\  -\cos\theta\sin\theta\cos2\theta\left(m_1e^{i\phi}-m_2e^{-i\phi}\right) & -2\sin^2\theta\cos^2\theta\left(m_1e^{2i\phi}-m_2\right) \end{array}\right).
\end{eqnarray}
All Yukawa couplings are invariant under $\theta\to\theta+\pi$ since all coefficients are in fact functions of $2\theta$.\footnote{Remember $2\sin^2\theta=1-\cos 2\theta$ and $2\cos^2\theta=1+\cos 2\theta$.} Furthermore, all Yukawa couplings are invariant under $\theta\to-\theta$ or $\phi\to\phi+\pi$ as long as one redefines the relative sign between the $\nu_1$ and $\nu_2$ fields. These are exactly the ``symmetries'' explored in \cite{deGouvea:2008nm}.

It is interesting to ask what happens in some well known cases. If lepton number is conserved, $g_{ij}=y_{ij}=0, \forall i,j$, and the two Weyl fermions merge into one Dirac neutrino. In this case, the parameters are as follows: $m_1=m_2=m$, $\theta=\pi/4$, $\phi=\pi/2$ and one can quickly check that, indeed,  $g_{ij}=y_{ij}=0, \forall i,j$. On the other hand, $\lambda^{\nu}_{ij}=0$ when $i\neq j$, while the diagonal elements $|\lambda^{\nu}_{11}|=|\lambda^{\nu}_{22}|=m/v$, as expected of the Yukawa coupling of a Dirac fermion in the mass basis.

If there are no Higgs triplets, $g=0$ and we are faced with the familiar type-I seesaw Lagrangian \cite{SeeSaw}. $g=0$ implies a non trivial relationship between $m_1,m_2,$ and $U$, which translates into $m_1\cos^2\theta+e^{-2i\phi} m_2\sin^2\theta=0$ or $\phi=\pi/2$ and $\tan^2\theta=m_1/m_2$. In the popular limit $m_2\gg m_1$, we recover the well known results: $\nu_1\sim\nu_e$, $\nu_2\sim\nu_s$, $y_{22}\sim m_2/s\gg y_{12,11}$, and $|\lambda^{\nu}_{12}|\sim \theta m_2/v\gg\lambda^{\nu}_{11,22}$. The last statement is the well-known seesaw relation: $(\lambda^{\nu}v)^2\sim m_1m_2$, or $m_1\sim(\lambda^{\nu}v)^2/m_2$ \cite{SeeSaw}.

The couplings $g,y,\lambda$ determine the strength of processes involving the propagating Higgs boson fields. These are, in general, linear combinations of the degrees of freedom in $H$, $T$, and $S$. For example, if $T$ and $S$ mixing with $H$ can be ignored, the off-diagonal Standard Model Higgs boson coupling to neutrinos can be written as 
\begin{equation}\label{1+1 type-I Higgs decay}
    -\frac{1}{2 v}\sin4\theta e^{i \phi } \left(e^{-2 i \phi } m_1-m_2\right)h^0\nu_1^{\dagger}
    \nu_2^{\dagger} + H.c.,
\end{equation}
and the matrix element squared for $h^0\to\nu_1\nu_2$ (assuming it is kinematically allowed) is 
\begin{equation}\label{type-I seesaw Higgas decay M^2}
    |i \mathcal{M}|^2=\frac{1}{2 v^2}\sin^24\theta(m_1^2+m_2^2-2m_1m_2\cos2\phi)\times(m^2_h-m^2_1-m^2_2)+
    \frac{1}{2v^2}\sin^24\theta((m_1^2+m_2^2)\cos2\phi-2m_1m_2)\times(2m_1m_2).
\end{equation}
The partial width $\Gamma(h^0\to\nu_1\nu_2)$ is in general nonzero and uniquely determined by the neutrino masses and the elements of the neutrino mixing matrix (including the Majorana phase $\phi$).

\begin{figure}
\includegraphics[width=0.4\textwidth]{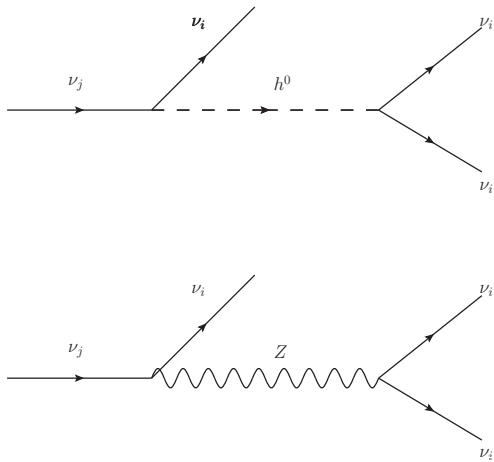}
\caption{Three level diagrams that mediate the decay of a neutrino mass eigenstate $\nu_j$ into three neutrino mass eigenstates $\nu_i$.}
\label{fig:onetothree decay}
\end{figure}

Other processes mediated by the Yukawa interactions include some  neutrino decays. For example, the process $\nu_2\to\nu_1\nu_1\nu_1$ (assuming it is kinematically allowed) is mediated by both Higgs and $Z$-boson exchanged, as depicted in Fig.~\ref{fig:onetothree decay}. In practice, given the minute size of the active neutrino masses, the $Z$-boson exchange dominates in most cases of interest. On the other hand, the $\nu_2\to\nu_1\gamma$ decay, which happens at the one-loop level, can be interpreted as a pure electroweak process if $T$ and $S$ are not available and all the charged Higgs-scalar degrees of freedom get ``eaten'' by the massive gauge bosons.  

\subsection{One Active, Two Sterile}
In the case of only one active ($\nu_e$) and two sterile neutrinos ($\nu_{s_1},\nu_{s_2}$) along with one charged lepton ($e$), the $3\times 3$ neutrino mixing matrix $U$ is parameterized by two mixing angles $\theta_{12},\theta_{13}$ and two Majorana phases $\phi_1,\phi_2$. 

The small number of mixing parameters is a consequence of the fact that one is free to rotate in the $\nu_{s_1},\nu_{s_2}$ space without impacting the weak interactions, which only depend on $U_{e1},U_{e2},U_{e3}$. This can be verified explicitly if we choose to parameterize $U$ in the usual way \cite{Amsler:2008zzb},
\begin{equation}
U=R^{23}(\theta_{23})R^{13}(\theta_{13},\delta)R^{12}(\theta_{12})\cdot\left(\begin{array}{ccc}1&0&0 \\ 0 & e^{i\phi_1} & 0 \\ 0 & 0 & e^{i\phi_2}\end{array}\right),~{\rm and}~m^{\nu}_{\rm Diag}=\left(\begin{array}{ccc}m_1&0&0 \\ 0 & m_2 & 0 \\ 0 & 0 & m_3\end{array}\right),
\label{U_2+1}
\end{equation} 
where 
\begin{equation}
R^{23}(\theta)=\left(\begin{array}{ccc}1&0&0 \\ 0 & \cos\theta & \sin\theta \\ 0 & -\sin\theta & \cos\theta \end{array}\right), 
R^{12}(\theta)=\left(\begin{array}{ccc}\cos\theta&\sin\theta&0 \\ -\sin\theta & \cos\theta & 0 \\ 0 & 0 & 1\end{array}\right),
R^{13}(\theta,\delta)=\left(\begin{array}{ccc}\cos\theta&0&\sin\theta e^{i\delta} \\ 0 & 1 & 0 \\ -\sin\theta e^{-i\delta} & 0 & \cos\theta \end{array}\right).
\label{R_ij}
\end{equation}
Complete expressions for the Yukawa couplings are on the long side and are not presented here. We do, however, invite readers to check that the ``symmetries'' identified in \cite{deGouvea:2008nm} are indeed present in this case. Here we compute the intermediate products
\begin{eqnarray}
U^{\dagger}\left(\begin{array}{ccc} 1 & 0 & 0 \\ 0 & 0 & 0 \\ 0 & 0 & 0 \end{array} \right)U &=& \left(\begin{array}{ccc} \cos^2\theta_{12}\cos^2\theta_{13} & \frac{1}{2}e^{i\phi_1}\sin2\theta_{12}\cos^2\theta_{13} & \frac{1}{2}e^{i(\phi_2+\delta)}\cos\theta_{12}\sin2\theta_{13} \\ \frac{1}{2}e^{-i\phi_1}\sin2\theta_{12}\cos^2\theta_{13}  & \sin^2\theta_{12}\cos^2\theta_{13} & \frac{1}{2}e^{i(-\phi_1+\phi_2+\delta)}\sin\theta_{12}\sin2\theta_{13} \\  \frac{1}{2}e^{-i(\phi_2+\delta)}\cos\theta_{12}\sin2\theta_{13} & \frac{1}{2}e^{i(\phi_1-\phi_2-\delta)}\sin\theta_{12}\sin2\theta_{13} & \sin^2\theta_{13} \end{array}\right), \nonumber \\
 U^{\dagger}\left(\begin{array}{ccc} 0 & 0 & 0 \\ 0 & 1 & 0 \\ 0 & 0 & 1 \end{array} \right)U &=& \left(\begin{array}{ccc} 
\sin^2\theta_{12}+\cos^2\theta_{12}\sin^2\theta_{13} & -\frac{1}{2}e^{i\phi_1}\sin2\theta_{12}\cos^2\theta_{13} & -\frac{1}{2}e^{i(\phi_2+\delta)}\cos\theta_{12}\sin2\theta_{13} \\ -\frac{1}{2}e^{-i\phi_1}\sin2\theta_{12}\cos^2\theta_{13} & \cos^2\theta_{12}+\sin^2\theta_{12}\sin^2\theta_{13} & -\frac{1}{2}e^{i(-\phi_1+\phi_2+\delta)}\sin\theta_{12}\sin2\theta_{13} \\ -\frac{1}{2}e^{-i(\phi_2+\delta)}\cos\theta_{12}\sin2\theta_{13} & -\frac{1}{2}e^{i(\phi_1-\phi_2-\delta)}\sin\theta_{12}\sin2\theta_{13} & \cos^2\theta_{13}
 \end{array} \right), \nonumber \\ \label{eq:1+2}
\end{eqnarray}
and note that the $g,y,\lambda^{\nu}$ depend on these matrices (plus their transposes) and $m^{\nu}_{\rm Diag}$. It is easy to see that, since $U$ is unitary, the two equations in  Eqs.~(\ref{eq:1+2}) add up to the identity matrix. All mixing angle dependency is captured in Eqs.~(\ref{eq:1+2}). As advertised, the unphysical ``sterile'' mixing angle $\theta_{23}$ is nowhere to be found, along with the unphysical Dirac phase $\delta$, which can be defined away via $\phi_2\to\phi_2-\delta$. In summary, all Yukawa couplings can be uniquely determined once one knows the three neutrino masses $m_1$, $m_2$, and $m_3$, the two Majorana phases $\phi_1,\phi_2$, and the two mixing angles $\theta_{12}$ and $\theta_{13}$. In the absence of right-handed neutrinos (or, say, if all $\lambda^{\nu}$ vanished) none of the mixing angles would be physical and the remaining Yukawa couplings ($g$, $y_{ij}$) would depend only on the neutrino mass eigenstate.

\subsection{Two Active, One Sterile}
In the case of two active ($\nu_e,\nu_{\mu}$) and one sterile neutrino ($\nu_s$) along with two charged leptons ($e,\mu$), the $3\times 3$ neutrino mixing matrix $U$ is parameterized by three mixing angles $\theta_{12},\theta_{13},\theta_{23}$, two Majorana phases $\phi_1,\phi_2$, and one Dirac phase $\delta$. Here, there are no redundant sterile neutrino field redefinitions that allow one to reduce the number of mixing parameters as far as the weak interactions are concerned. 

Here we choose to parameterize $U$ in a slightly unorthodox way, namely,
\begin{equation}
U=R^{12}(\theta_{12})R^{13}(\theta_{13},\delta)R^{23}(\theta_{23})\cdot\left(\begin{array}{ccc}1&0&0 \\ 0 & e^{i\phi_1} & 0 \\ 0 & 0 & e^{i\phi_2}\end{array}\right),~{\rm and}~m^{\nu}_{\rm Diag}=\left(\begin{array}{ccc}m_1&0&0 \\ 0 & m_2 & 0 \\ 0 & 0 & m_3\end{array}\right), \label{U_1+2}
\end{equation}
where the matrices $R^{12,13,23}$ are defined in Eq.~(\ref{R_ij}). Eq.~(\ref{U_1+2}) is to be contrasted to Eq.~(\ref{U_2+1}). As in the previous subsection, complete expressions for the Yukawa couplings are long and are not included. As before, we quote the results for
\begin{eqnarray}
U^{\dagger}\left(\begin{array}{ccc} 0 & 0 & 0 \\ 0 & 0 & 0 \\ 0 & 0 & 1 \end{array} \right)U &=& \left(\begin{array}{ccc} 
\sin^2\theta_{13} & \frac{1}{2}e^{i(\phi_1+\delta)}\sin2\theta_{13}\sin\theta_{23} & -\frac{1}{2}e^{i(\phi_2+\delta)}\sin2\theta_{13}\cos\theta_{23} \\
\frac{1}{2}e^{-i(\phi_1+\delta)}\sin2\theta_{13}\sin\theta_{23} & \cos^\theta_{13}\sin^2\theta_{23} & -\frac{1}{2}e^{-i(\phi_1-\phi_2)}\cos^2\theta_{13}\sin2\theta_{23} \\
 -\frac{1}{2}e^{-i(\phi_2+\delta)}\sin2\theta_{13}\cos\theta_{23} & -\frac{1}{2}e^{i(\phi_1-\phi_2)}\cos^2\theta_{13}\sin2\theta_{23} & \cos^2\theta_{13}\cos^2\theta_{23}
\end{array}\right), \nonumber \\
U^{\dagger}\left(\begin{array}{ccc} 1 & 0 & 0 \\ 0 & 1 & 0 \\ 0 & 0 & 0 \end{array} \right)U &=&  \left(\begin{array}{ccc} 1 & 0 & 0 \\ 0 & 1 & 0 \\ 0 & 0 & 1 \end{array} \right)- U^{\dagger}\left(\begin{array}{ccc} 0 & 0 & 0 \\ 0 & 0 & 0 \\ 0 & 0 & 1 \end{array} \right)U.
\end{eqnarray} 
From the expressions above, it is clear that, in the mass basis, none of the Yukawa couplings $g,y,\lambda^{\nu}$ depend on the ``purely active'' mixing angle $\theta_{12}$ or the physical Dirac phase $\delta$ which can be defined away via $\phi_1\to\phi_1-\delta$, $\phi_2\to\phi_2-\delta$. This is in spite of the fact that the weak interactions do depend on both $\theta_{12}$ and $\delta$. In summary, all Yukawa couplings are uniquely determined once one knows the three neutrino masses $m_1$, $m_2$, and $m_3$, the two Majorana phases $\phi_1,\phi_2$, and only two of the three mixing angles, $\theta_{13}$ and $\theta_{23}$. Yukawa interactions do not depend on $\theta_{12}$ or $\delta$. In the absence of right-handed neutrinos (or, say, if all $\lambda^{\nu}$ vanished) only the mixing angle $\theta_{12}$ would be physical (along with the Majorana phase $\phi_1$) and the remaining Yukawa couplings would depend only on the neutrino mass eigenvalues.

\subsection{More than One Active, More than One Sterile}

In the case of several active and sterile neutrino states, it is possible to express all Yukawa couplings in terms of the neutrino mass eigenvalues, all the Majorana phases, and only a subset of the physical mixing angles and Dirac phases. This is a consequence of the equivalence, as far as the Yukawa interactions are concerned, described in Eq.~(\ref{eq:Omega_a}). As a concrete example, consider two active and two sterile neutrinos. The physical mixing parameters are: five mixing angles, two Dirac phases and three Majorana phases. Of those, the Yukawa interactions do {\sl not} depend on one  mixing angle and one Dirac phase. If we define $U$ as
\begin{eqnarray}\label{2+2 type-I MNS}
    U &=& 
       U^DR^{23}R^{14}R^{13}R^{24}P^{\phi}=
    \left(\begin{array}{cc} \Omega^a_{2\times 2} & 0_{2\times 2} \\ 0_{2\times 2} & \Omega_{2\times 2}\end{array}\right)
      \left(
        \begin{array}{cccc}
          1 & 0 & 0 & 0 \\
          0 & \cos\theta_{23} & \sin\theta_{23} & 0 \\
          0 & -\sin\theta_{23} & \cos\theta_{23} & 0 \\
          0 & 0 & 0 & 1 \\
        \end{array}
      \right)
      \left(
        \begin{array}{cccc}
          \cos\theta_{14} & 0 & 0 & \sin\theta_{14} e^{i\delta} \\
          0 & 1 & 0 & 0 \\
          0 & 0 & 1 & 0 \\
          -\sin\theta_{14} e^{-i\delta} & 0 & 0 & \cos\theta_{14} \\
        \end{array}
      \right)  \nonumber \\
      &\cdot & \left(
        \begin{array}{cccc}
          \cos\theta_{13} & 0 & \sin\theta_{13} & 0 \\
          0 & 1 & 0 & 0 \\
          -\sin\theta_{13} & 0 & \cos\theta_{13} & 0 \\
          0 & 0 & 0 & 1 \\
        \end{array}
      \right)
       \left(
        \begin{array}{cccc}
          1 & 0 & 0 & 0 \\
          0 & \cos\theta_{24} & 0 & \sin\theta_{24} \\
          0 & 0 & 1 & 0 \\
          0 & -\sin\theta_{24} & 0 & \cos\theta_{24} \\
        \end{array}
      \right)
      \left(
         \begin{array}{cccc}
           1 & 0 & 0 & 0 \\
           0 & e^{i\phi_1} & 0 & 0 \\
           0 & 0 & e^{i\phi_2} & 0 \\
           0 & 0 & 0 & e^{i\phi_3} \\
         \end{array}
       \right), 
       \label{U_2+2}
\end{eqnarray}
the unphysical parameters in $\Omega_{2\times 2}$ (one angle and one phase) and the physical parameters in $\Omega^a_{2\times 2}$ will be explicitly absent from all $y,g,\lambda^{\nu}$ in the mass basis.

\setcounter{equation}{0}
\section{Comment on The Casas--Ibarra Parameterization}
\label{sec:Casas}

Of particular interest is the so-called type-I seesaw Lagrangian. In the language developed in Sec.~\ref{sec:Lagrangian}, this corresponds to $g_{ij}=0, \forall i,j$. We also constrain the discussion to the case where $S$ is not a propagating field but instead lepton number is explicitly broken by Majorana mass parameters for the gauge-singlet fields, $M^R$. In this case
\begin{equation}
m^{\nu} = \left(\begin{array}{cc} 0  & \lambda^{\nu}v \\ (\lambda^{\nu}v)^T & M^R \\ \end{array} \right).
\end{equation}
In the limit that the elements of $M^R$ are much larger than those of $\lambda^{\nu}v$, one can diagonalize $m^{\nu}$ perturbatively. Following the notation in \cite{de Gouvea:2007uz}, 
\begin{eqnarray}
 U &=& \left(
        \begin{array}{cc}
          V & \Theta \\
          -\Theta^{\dagger}V & 1 \\
        \end{array}
      \right),\label{eq:U_CI} \\
  \Theta &=& V\sqrt{\mbox{diag}(m_1,m_2,m_3)}R^{\dagger}[\mbox{diag}(m_4,m_5,\ldots)]^{-1/2}, \label{eq:Theta}
\end{eqnarray}
where $V$ is a unitary $3\times 3$ matrix, $m_1,m_2,m_3$ are the light neutrino masses and $m_4,m_5,\ldots$ are the heavy ones. $R_{N\times 3}$ is a complex orthogonal matrix $RR^T=1$ and we assume $N\le 3$.\footnote{In the case $N>3$, it is easy to write an equivalent version of Eq.~(\ref{eq:Theta}) in terms of a different complex orthogonal matrix.} This very useful parameterization is due to Casas and Ibarra \cite{Casas:2001sr} and aims at separating the ``light'' mixing parameters, contained in $V$, from the ``heavy'' ones, contained in $R$. In this case, one chooses all ``light'' parameters to be independent along with the heavy mass eigenstates so that the ``heavy--light'' mixing parameters are not generic but constrained due to the $g_{ij}=0, \forall i,j$ constraint of the type-I seesaw Lagrangian. The more general case $g_{\ij}\neq 0$ but $g_{ij}u\ll m_4,m_5,\ldots$ has also been discussed in the literature \cite{Akhmedov:2008tb} but  here we will stick to the type-I case presented above.

A quick parameter counting is in order. Assuming $N=3$ for concreteness, $m^{\nu}$ contains, after all field redefinitions, 18 real independent parameters. After diagonalization, we are left with 3 light masses, 3 heavy masses (6 real parameters), $V$ is parameterized by three mixing angles, two Majorana phases and one Dirac phase (6 real parameters), and the $3\times 3$ complex orthogonal matrix is parameterized by three complex ``angles,'' which translate into 6 real parameters. Not surprisingly, the total is 18, as required. Note that this is much less than the number of parameters for a generic $6\times 6$ $m^{\nu}$ which requires 6 masses and 24 real parameters in the mixing matrix $U$. The constraint $g_{ij}=0, \forall i,j$ imposes 12 nontrivial relations among the 30 parameters, reducing the parameter space to 18.  

In the mass basis, the Yukawa interactions are given by, at leading order in $m_{\rm light}/m_{\rm heavy}$:
\begin{equation}
\lambda^{\nu} = \frac{1}{v}\left(\begin{array}{cc}0_{3\times 3} & \sqrt{\mbox{diag}\left(m_1,m_2,m_3\right)}R^T\sqrt{\mbox{diag}\left(m_4,m_5,\ldots\right)} \\  \sqrt{\mbox{diag}\left(m_4,m_5,\ldots\right)}R\sqrt{\mbox{diag}\left(m_1,m_2,m_3\right)} & 0_{N\times N}\end{array}\right) + {\cal O}\left(\frac{m_{\rm light}}{v}\right),
\end{equation}
where $m_{\rm light}=m_1,m_2,m_3$, $m_{\rm heavy}=m_4,m_5,\ldots$. As expected, the Yukawa couplings in the mass basis are independent of the ``active'' mixing angles in $V$ and only depend on the ``active--sterile'' mixing angles. In the Casas--Ibarra parameterization, these are parameterized by the neutrino masses and $R$. On the other hand, in the flavor basis, the entries of the $3\times N$ Yukawa matrix are given by (see, for example, \cite{de Gouvea:2007uz}),
\begin{equation}
\lambda^{\nu}_{\alpha i}=\frac{1}{v}\left[V^*\sqrt{\mbox{diag}\left(m_1,m_2,m_3\right)}R^T\sqrt{\mbox{diag}\left(m_4,m_5,\ldots\right)}\right]_{\alpha i}. 
\label{eq:lambda_flavor}
\end{equation}
Here $\alpha$ runs over the active neutrino flavors (say, $\nu_{e},\nu_{\mu},\nu_{\tau}$) while $i$ runs over the sterile neutrino ``flavors'' (from 1 to $N$).

Eq.~(\ref{eq:U_CI}), since it treats active and sterile degrees of freedom differently, casts $U$ in an unusual form (compared, say, to Eqs.~(\ref{U_2+1},\ref{U_1+2},\ref{U_2+2})). For example, there are no ``Majorana Phases'' associated (as discussed in \cite{deGouvea:2002gf}) to the heavy masses $m_4,m_5,\ldots$. Instead, these are included in $R$. In addition, the parameters in $R$ are complex ``mixing angles'' such that the magnitudes of its elements are not bound (unlike the elements of $U$, whose magnitudes are constrained to be less than 1).\footnote{The seesaw approximation requires the elements of $\Theta\ll 1$ so, in practice, the elements of $R$ are bound to be smaller, in magitude, than $\sqrt{m_{\rm heavy}/m_{\rm light}}$.} It is, hence, difficult and perhaps not illuminating to discuss the range of parameters within $R$ that allow one to cover all physically distinguishable possibilities. $V$, on the other hand, is the familiar active leptonic mixing matrix and, as such, we would like to make sure it can be parameterized in the standard way \cite{deGouvea:2008nm}: $\theta_{12,13,23}\in[0,\pi/2]$, $\delta\in[0,2\pi\}$, $\phi_{1,2}\in[0,\pi]$. Fortunately, Eq.~(\ref{eq:U_CI}) allows for that, as long as one also properly defines the physical range for the parameters of $R$.

As already discussed, $V(\theta,\phi,\delta)=P_\alpha V(\theta',\phi',\delta')P_i$ where $P$ are diagonal matrices with diagonal elements equal to $\pm1$. If $P_i$ ($i=1,2,3$) and $P_{\alpha}$ ($\alpha=e,\mu,\tau$) can be ``rotated away'' by properly redefining the charged lepton and neutrino fields, than one can reduce the parameter space down to the one listed above. Following this strategy here one gets:
\begin{eqnarray}
  U(\theta,\phi,\delta) &=& \left(
                  \begin{array}{cc}
                    V(\theta,\phi,\delta) & V(\theta,\phi,\delta)m^{1/2}R^{\dagger}M^{-1/2} \\
                    -M^{-1/2}R m^{1/2} & 1_{3\times 3} \\
                  \end{array}
                \right),
   \\
   &=& \left(
                  \begin{array}{cc}
                    P_\alpha V(\theta',\phi',\delta')P_i & P_{\alpha}V(\theta',\phi',\delta')P_im^{1/2}R^{\dagger}M^{-1/2} \\
                    -M^{-1/2}R m^{1/2} & 1_{3\times 3} \\
                  \end{array}
                \right), \notag
    \\
   &=&
   \left(
     \begin{array}{cc}
       P_{\alpha} & 0 \\
       0 & 1_{3\times 3} \\
     \end{array}
   \right)
  \left(
                  \begin{array}{cc}
                    V(\theta',\phi',\delta')P_i & V(\theta',\phi',\delta')m^{1/2}P_iR^{\dagger}M^{-1/2} \\
                    -M^{-1/2}R m^{1/2} & 1_{3\times 3} \\
                  \end{array}
                \right),
   \notag \\
   &=&
      \left(
     \begin{array}{cc}
       P_{\alpha} & 0 \\
       0 & 1_{3\times 3} \\
     \end{array}
   \right)
  \left(
                  \begin{array}{cc}
                    V(\theta',\phi',\delta') & V(\theta',\phi',\delta')m^{1/2}P_iR^{\dagger}M^{-1/2} \\
                    -M^{-1/2}R m^{1/2}P_i & 1_{3\times 3} \\
                  \end{array}
                \right)
                \left(
                  \begin{array}{cc}
                    P_i & 0 \\
                    0 & 1_{3\times 3} \\
                  \end{array}
                \right),
    \notag\\
   &=& \left(
     \begin{array}{cc}
       P_{\alpha} & 0 \\
       0 & 1_{3\times 3} \\
     \end{array}
   \right)
  \left(
                  \begin{array}{cc}
                    V(\theta',\phi',\delta') & V(\theta',\phi',\delta')m^{1/2}P_iR^{\dagger}M^{-1/2} \\
                    -M^{-1/2}RP_i m^{1/2} & 1_{3\times 3} \\
                  \end{array}
                \right)
                \left(
                  \begin{array}{cc}
                    P_i & 0 \\
                    0 & 1_{3\times 3} \\
                  \end{array}
                \right),\notag
\end{eqnarray}
where $m^{1/2}\equiv\sqrt{\mbox{diag}(m_1,m_2,m_3)}$ and $M^{-1/2}=\left({\mbox{diag}(m_4,m_5,m_6)}\right)^{-1/2}$ and we are constraining the discussion to $N=3$. While $P_{\alpha}$ and $P_i$ can be absorbed by redefining the charged-lepton and neutrino mass eigenstates, one ends up equating $\theta,\phi,\delta$ with $\theta',\phi',\delta'$ only if, at the same time, one also changes $R$ into $R'=P_iR$. This indicates that if one chooses the range of parameters in the active leptonic mixing matrix in the standard fashion, one must be sure to properly parameterize $R$ and explore the full physical range of its parameters. This is similar to the case of $\theta_{13}$ and $\delta$ when $N=0$ (no sterile neutrinos). One can either choose $\theta_{13}\in[0,\pi/2]$ and $\delta\in[0,2\pi\}$ or $\theta_{13}\in[-\pi/2,\pi/2]$ and  $\delta\in[0,\pi\}$, as discussed in \cite{deGouvea:2008nm,Latimer:2004gz,Fogli:2005cq}.

It is instructive to consider a concrete example. If there are two active and two sterile neutrinos, one can write, using the matrices defined in Eq.~(\ref{U_2+2}),
\begin{equation}
U=R^{23}R^{14}R^{13}R^{24} 
\left(\begin{array}{cccc}\cos\theta_{12} & \sin\theta_{12} & 0 & 0 \\ -\sin\theta_{12} & \cos\theta_{12} & 0 & 0 \\ 0 & 0 & 1 & 0 \\ 0 & 0 & 0 & 1 \end{array}\right) P^{\phi}.
\end{equation}
We further multiply $U$ from the left by diag$\left(1,1,e^{-i\phi_2},e^{-i\phi_3}\right)$, which leaves the rest of the Lagrangian unchanged upon simple field redefinitions. In the limit $\theta_{13},\theta_{14},\theta_{23},\theta_{24}\ll 1$, $U$ is given by Eq.~(\ref{eq:U_CI}), where
\begin{equation}
V=\left(\begin{array}{cc} \cos\theta_{12} & \sin\theta_{12} \\ -\sin\theta_{12} & \cos\theta_{12} \end{array}\right)\left(\begin{array}{cc} 1 & 0 \\ 0 & e^{i\phi_1}\end{array}\right), ~~~
\Theta=\left(\begin{array}{cc} \theta_{13}e^{i\phi_2} & \theta_{14}e^{i(\delta+\phi_3)} \\ \theta_{23}e^{i\phi_2} & \theta_{24}e^{i\phi_3}\end{array}\right).
\end{equation}
In the type-I seesaw, not all of these parameters are independent. Using the Casas--Ibarra parameterization,
\begin{equation}
\Theta = \left(\begin{array}{cc}
\sqrt{\frac{m_1}{m_3}}\cos\theta_{12}C \mp\sqrt{\frac{m_2}{m_3}}\sin\theta_{12}e^{i\phi_1}S & \sqrt{\frac{m_1}{m_4}}\cos\theta_{12}S \pm \sqrt{\frac{m_2}{m_4}}\sin\theta_{12}e^{i\phi_1}C \\ 
- \sqrt{\frac{m_1}{m_3}}\sin\theta_{12}C\mp\sqrt{\frac{m_2}{m_3}}\cos\theta_{12}e^{i\phi_1}S & -\sqrt{\frac{m_1}{m_4}}\sin\theta_{12}S \pm \sqrt{\frac{m_2}{m_4}}\cos\theta_{12}e^{i\phi_1}C
\end{array}\right),
\end{equation}
where $C$ and $S$, are the complex parameters in $R$, 
\begin{equation}
R^{\dagger}=\left(\begin{array}{cc} C & S \\ \mp S & \pm C\end{array} \right), \label{eq:R_2x2}
\end{equation}
$C^2+S^2=1$. The $\pm$ sign is worthy of a few sentences. Since $R$ is an orthogonal matrix, its determinant can be either $+1$ or $-1$. There is no way of parameterizing $R$ as a single function of continuous complex parameters and cover both determinants so the two discrete choices for the sign need to be included. There are, however, choices for the ranges of the other parameters that render both possible determinants of $R$ physically equivalent. A simple one is to allow $\phi_1\in[0,2\pi\}$ instead of $\phi_1\in[0,\pi]$, as is typically the case \cite{deGouvea:2008nm}. This choice was made, for example, in \cite{Molinaro:2008rg,Gavela:2009cd}. It is easy to see that $+\to -$ combined with $\phi_1\to\phi_1+\pi$ leaves $\Theta$ unchanged. There are potentially other choices. Since we haven't specified the physical range of $S$ and $C$, it is easy to note that $+\to -$ and $C\to-C$ or $S\to S$ lead to an overall sign change to one of the two columns of $\Theta$. Such a change is not physically observable.  

\begin{figure}
\includegraphics[width=0.5\textwidth]{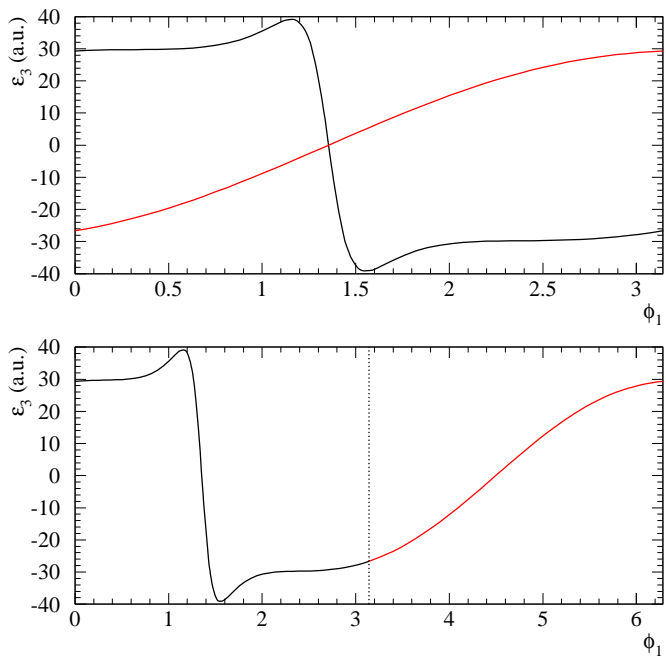}
\caption{$\epsilon_3$, as defined in Eq.~(\ref{eq:e3}), as a function of the Majorana phase $\phi_1$ for $m_2=2m_1$, $m_3=100m_1$, $m_4=1000m_1$, $\theta_2=\pi/5$, $C=\sqrt{2}e^{-i\pi/6}$, and $S=3^{1/4}e^{-i3\pi/4}$, in arbitrary units. In the top panel, det$(R)=+1$ (black curve) and det$(R)=-1$ (red [grey] curve) clearly lead to different values of $\epsilon_3$. In the bottom panel we fix det$(R)=+1$ and double the $\phi_1$ range. The vertical dotted line corresponds to $\phi_1=\pi$. It is clear that the imagine of $\phi_1\in[\pi,2\pi]$ is the same as the imagine of the det$(R)=-1$, $\phi_1\in[0,\pi]$ points in the top figure. }
\label{fig:lepto}
\end{figure}

We now compute the lepton asymmetry $\epsilon_3$ generated in the decay of  a $\nu_3$. $\epsilon_3$ is related to how large a baryon asymmetry the universe acquires in the case of thermal leptogenesis \cite{Fukugita:1986hr,lepto_review,lepto_rev_2}. 
\begin{equation}
    \epsilon_3=\frac{\Gamma(\nu_3\to HL)-\Gamma(\nu_3\to H^*\bar{L})}{\Gamma(\nu_3\to HL)+\Gamma(\nu_3\to H^*\bar{L})}
    \simeq \frac{3}{8\pi}\frac{\text{Im}\{(\lambda^{\nu}(\lambda^{\nu})^{\dagger})^2_{12}\}}{(\lambda^{\nu}(\lambda^{\nu})^{\dagger})_{11}} \frac{m_3}{m_4},\label{eq:e3}
\end{equation}
where we assume $m_1,m_2\ll m_3\ll m_4$ (see, for example, \cite{lepto_rev_2}) and $\lambda^{\nu}$ is expressed in the flavor basis, Eq.~(\ref{eq:lambda_flavor}). In Fig.~\ref{fig:lepto} we depict $\epsilon_3$ as a function of the Majorana phase $\phi_1$ for $m_2=2m_1$, $m_3=100m_1$, $m_4=1000m_1$, $\theta_2=\pi/5$, $C=\sqrt{2}e^{-i\pi/6}$ and $S=3^{1/4}e^{-i3\pi/4}$ (note $C^2+S^2=1$) and the two distinct values for the determinant of $R=\pm1$, see Eq.~(\ref{eq:R_2x2}). All possible distinct values of $\epsilon_3$ can be obtained (assuming all other parameters fixed!) if one allows for both values of det$(R)$ and varies $\phi_1\in[0,\pi\}$ (Fig~\ref{fig:lepto}-top) or if one fixes det$(R)$ and varies $\phi_1\in[0,2\pi\}$ (Fig~\ref{fig:lepto}-bottom). The exact same effect can be obtained by changing the sign of either $C$ or $S$. For example, by keeping all other parameters fixed, including det$(R)$, but allowing for $C=\sqrt{2}e^{-i\pi/6}$ {\sl and} $C=-\sqrt{2}e^{-i\pi/6}$ one also obtains the plot depicted in Fig~\ref{fig:lepto}-top (black line for $C=\sqrt{2}e^{-i\pi/6}$, red [grey] for $C=-\sqrt{2}e^{-i\pi/6}$).

\section{Conclusions} 
\label{sec:Conclusion}

Active neutrino masses, like all Standard Model charged-fermion masses, only arise after electroweak symmetry breaking. This is true regardless of the nature of the neutrinos or the mechanism responsible for their finite masses. If the mechanism of electroweak symmetry breaking can be captured by the vacuum expectation values of scalar fields, fermion masses arise due to the interactions between the fermions and these scalar fields. 

In the case of the charged-fermions in the  Standard Model, charged-fermions will couple to the propagating Higgs boson in a mass-diagonal way, and the magnitude of these couplings is proportional to the individual charged-fermion masses, as is well known. This is also the case of neutrinos if they are massive Dirac fermions. In this case, of course, the Standard Model Lagrangian is augmented to include at least two Standard Model gauge singlet Weyl fermions, the so-called right-handed neutrinos. 

If the neutrinos are Majorana fermions and if right-handed neutrinos are propagating degrees of freedom, the couplings of neutrinos to the Higgs sector will depend not only on the neutrino masses but also on a subset of the mixing parameters that define the full neutrino mixing matrix. Furthermore, in the neutrino mass-eigenstate basis, the couplings are not necessarily diagonal. We discussed in detail how to properly parameterize the couplings of the neutrino--Higgs sector. We find that neutrino Yukawa couplings are well-defined functions of all the neutrino mass eigenvalues and the ``active--sterile'' mixing parameters in the full neutrino mixing matrix. Furthermore, we can still choose to constrain the ranges of the parameters that define the elements of the mixing matrix to the ones that are required in order to properly describe the weak interactions (see, for example, \cite{deGouvea:2008nm,Latimer:2004hd,Jenkins:2007ip}). This means if we choose all neutrino masses to be positive, allow all mixing angles to lie in the first quadrant ($\theta\in[0,\pi/2]$), and allow all Majorana phases to lie in the first two quadrants ($\phi\in[0,\pi]$) we will properly parameterize all physically distinguishable outcomes for processes involving interactions between all leptons (including right-handed neutrinos) and gauge bosons or Higgs bosons, as long as all Dirac phases are allowed to vary within the entire unit circle ($\delta\in[0,2\pi\}$). This result applies not only to the lepton couplings to the Standard Model Higgs doublet but can be extended to include a Higgs boson $SU(2)_L$ triplet and a Higgs boson Standard Model singlet. This is true as long as all elements of the neutrino mass matrix are proportional to the vacuum expectation values of these different scalar fields.  

Unlike the quark mixing sector, many of the potentially physical parameters in the lepton mixing sector are completely unknown. We don't know whether there are right-handed neutrinos, Higgs-boson triplets, or Higgs-boson singlets, we don't know whether neutrinos are Majorana fermions, and we don't even know the magnitudes of the mostly-acitve neutrino masses. We only know that there are at least three neutrinos, and we have measured two mass-squared differences and the values of two mixing angles. We can infer that a third mixing angle exists, along with a Dirac phase, although both of them could be zero. 

Regardless of all the uncertainty, many of the hypothetical parameters, if they exist, play a pivotal role in potentially important processes, including leptogenesis, expectations of supersymmetric theories for charged-lepton flavor-violating processes, etc. For this reason we discussed relations between the parameterization discussed here and the Casas--Ibarra parameterization, and discussed the physical range of the parameters in the latter case. 

In conclusion, the issue of properly parameterizing both the gauge and Yukawa sectors and understanding how to access the entire allowed parameter space is particularly important in the neutrino sector. We also have empirical evidence that it is an issue that often leads to misunderstanding among even seasoned neutrino phenomenologists and to confusing statements in the neutrino literature. We hope our results will help clarify some of the confusion. 

Finally, we will emphasize some important provisos. The results presented here apply assuming that, other than the Yukawa interactions highlighted above, there are no new interactions involving active or sterile neutrinos. If for example, the right-handed neutrinos couple to some singlet fermion $S$ in a way that these new couplings $y'$ are not related to the right-handed neutrino Majorana mass parameters $M^R_{ij}$, the $y'$ couplings will not depend only on neutrino masses and mixing angles but will also depend on some extra parameters. We also assume that the new neutrino degrees of freedom are Standard Model gauge singlets. If one were to enlarge the particle content of the Standard Model by including fermionic $SU(2)_L$ triplets with zero hypercharge, nonzero neutrino masses would be generated (this is the type-III seesaw \cite{Foot:1988aq}), but the neutrino mixing space would be modified since fermion triplets also couple to the $W$-boson. 
  
%
%
%
%
%
%

\section*{Acknowledgments}
AdG thanks Serguey Petcov for interesting questions and enlightening discussions that ultimately led to the pursuit of this project.   
This work is sponsored in part by the US Department of Energy Contract DE-FG02-91ER40684.

 \end{document}